# A Scoping Review of Internal Migration and Left-behind Children's Wellbeing in China

Jinkai Li*

Department of Economics, Faculty of Economics and Business Administration, Ghent University, Ghent 9000, Belgium. jinkai.li@ugent.be (Jinkai Li)

*Correspondence: jinkai.li@ugent.be

**Abstract:** Children's well-being of immigrants is facing several challenges related to physical, mental, and educational risks, which may obstacle human capital accumulation and further development. In rural China, due to the restriction of Hukou registration system, nearly 9 million left-behind children (LBC) are in lack of parental care and supervision in 2020 when their parents internally migrate out for work. Through the systematic scoping review, this study provides a comprehensive literature summary and concludes the overall negative effects of parental migration on LBC's physical, mental (especially for left-behind girls), and educational outcomes (especially for left-behind boys). Noticeably, both parents' and mother's migration may exacerbate LBC's disadvantages. Furthermore, remittance from migrants and more family-level and social support may help mitigate the negative influence. Finally, we put forward theoretical and realistic implications which may shed light on potential research directions. Further studies, especially quantitative studies, are needed to conduct a longitudinal survey, combine the ongoing Hukou reform in China, and simultaneously focus on left-behind children and migrant children.

## 1. Introduction

Immigrants nearly account for 15% of the total population worldwide [1], in which internal migration is an important component. Within the context of global growing migration, children's well-being of immigrants may be vulnerable, which has been an important issue [2]. This is also closely related to the achievements of *Sustainable Development Goal 3* (Good health and Well-being) and *Sustainable Development Goal 10* (Reduced Inequalities) [3,4]. In China, there are an increasing number of internal migration from rural to urban areas or from small to large cities [5]. Due to the limitation of family wealth and social institutions, rural children are more likely to be left behind and vulnerable in hometowns by their parents [6]. There has been abundant literature demonstrating that parental companionship and supervision are critical for children's development in early childhood [7–9].

*Hukou*, the Chinese household registration system, offers a unique opportunity to study how parental (internal) migration affects left-behind children's (LBC) development. The *Hukou* system divides Chinese residents into two categories, agricultural *Hukou* and non-agricultural *Hukou*, depending on whether they are born in rural or urban areas [10]. Usually, if immigrants holding agricultural *Hukou* live in cities, they will not receive the equivalent public service as urban residents holding non-agricultural *Hukou*, such as access to housing property, education, and health service [11]. Consequently, if both parents migrate to cities, their children are forced to be left behind and taken care of by their grandparents or other family relatives in rural China [12].

It is of great importance to identify the causal effect of parental migration on LBC development at an early stage. First, over 61 million children were left behind by at least one of their parents in 2014. As the report of the Ministry of Civil Affairs of China, in 2020, there are still 9 million children left behind by both parents in rural China. Second, most LBC are in the early stage of development. According to the "first 1000 days" theory [13–15], studies with followed-up children have proved that early parental companionship and investment may have a profound impact on children's cognitive and non-cognitive development [16,17].

We mainly conduct the following research procedures. First, we conduct the scoping review within the Arksey and O'Malley framework and survey 33 empirical studies which estimate the overall effect of parental migration on LBC's physical, mental, and educational outcomes. Second, we clarify the complex mechanisms between parental migration and LBC's development, as well as the heterogeneity analysis. Last, we put forward relative implications and give conclusions.

This study may contribute to the literature in the following aspects. First, as we know, this study may be the first to conduct a systematic scoping review on LBC's physical, mental, and educational development simultaneously, which provides an overall and accurate analysis of LBC's well-being. Second, this study tries to conceptualize the theoretical framework for parental migration and LBC's well-being, simultaneously integrating mechanism analysis (the channel of remittance, parental practice, and social support) and heterogeneity analysis (the migration type and timing, parental income, and education, and LBC's gender).

The organization of this paper is as follows. Section 1 is an introduction. Section 2 gives a detailed explanation of the method of scoping review. Section 3 concludes the overall impact of parental migration on LBC's well-being from three aspects. In Section 4, this study provides a comprehensive mechanism and heterogeneity analysis. Section 5 includes conclusions and implications.

## 2. Method

The scoping review approach is an appropriate method to identify specific research questions or key research gaps [18,19]. To obtain accurate research conclusions, we follow the Arksey and O'Malley framework [20], which mainly includes five steps: (1) identify the research question; (2) identify the published studies which is closely related to the research question; (3) refine the study selection criteria; (4) conduct the summary of study characteristics; (5) summarize, report, and interpret the results. We provide the analysis results based on the PRISMA Scoping Review (PRISMA-ScR) checklist [21].

**Identify the research question**. We put forward the following research question: How does parental migration affect LBC's well-being? More specifically, based on three domains of children's well-being, we aim at identifying the impact of parental migration on LBC's physical, mental, and educational outcomes.

**Table 1.** Search terms

| | **Keywords** |
|---|---|
| **Physical outcomes** | Height, Height-for-age Z Score, Weight, Weight-for-age Z Score, Body mass index, Self-rated health, Stunting, Underweight, Wasting, Anemia, Blood pressure, Illness or chronic disease |
| **Mental outcomes** | Behavior problems index, Self-reported mental health index, Epidemiological depression score, Social anxiety index, Depression symptoms, Self-rated happiness and satisfaction, Big five personality index, Social-emotional evaluation |
| **Educational outcomes** | Educational aspiration, Accepted education level, Standardized English score, Verbal test score, Chinese/math test score, Final exam score, Cognitive score of Bayley scales of infant development, Self-reported academic performance |
| **Migration** | Parental migration, Parental absence, Internal migration, immigrants, rural-urban migration |
| **China** | rural China, Northwestern China, rural families of China |

**Identify relevant published studies.** The systematic study identification is carried out among four peer review journal databases (Science Direct, PubMed, Web of Science, and Scopus). The search period is from 1 November 2014 to 1 February 2022. The search terms consist of relative keyword combinations of Physical outcomes OR Mental outcomes OR Educational outcomes AND Migration AND China as shown in **Table 1** (Search terms).

**Refine the selection criteria.** The criteria for selecting literature are as follows. First, all research articles have been published in SCIE/SSCI indexed journals in the fields of health, education, psychology, migration, child development, and economics. Second, I focus on quantitative analysis of the impact of internal parental migration on LBC aged 0-18, which includes both causation and correlation analysis. Methods for causality inference mainly include FE, PSM, IV, and DID, whereas methods for correlation analysis mainly include linear or non-linear regressions. Third, I focus on outcomes that have been commonly used, which is possible to compare findings and verify whether findings are consistent in line with each other. Detailed descriptions of outcomes have been reported in **Table 1**. Fourth, we focus on LBC in rural China. Due to the restriction of *Hukou* household registration system in China, nearly 9 million left-behind children (LBC) are in lack of parental care and supervision in 2020 when their parents migrate out for work. The study flow chart is presented in **Figure 1**. Additionally, **the summary of study characteristics and detailed analyses** can be seen in Section 3 (**Table 2, 3, and 4**).

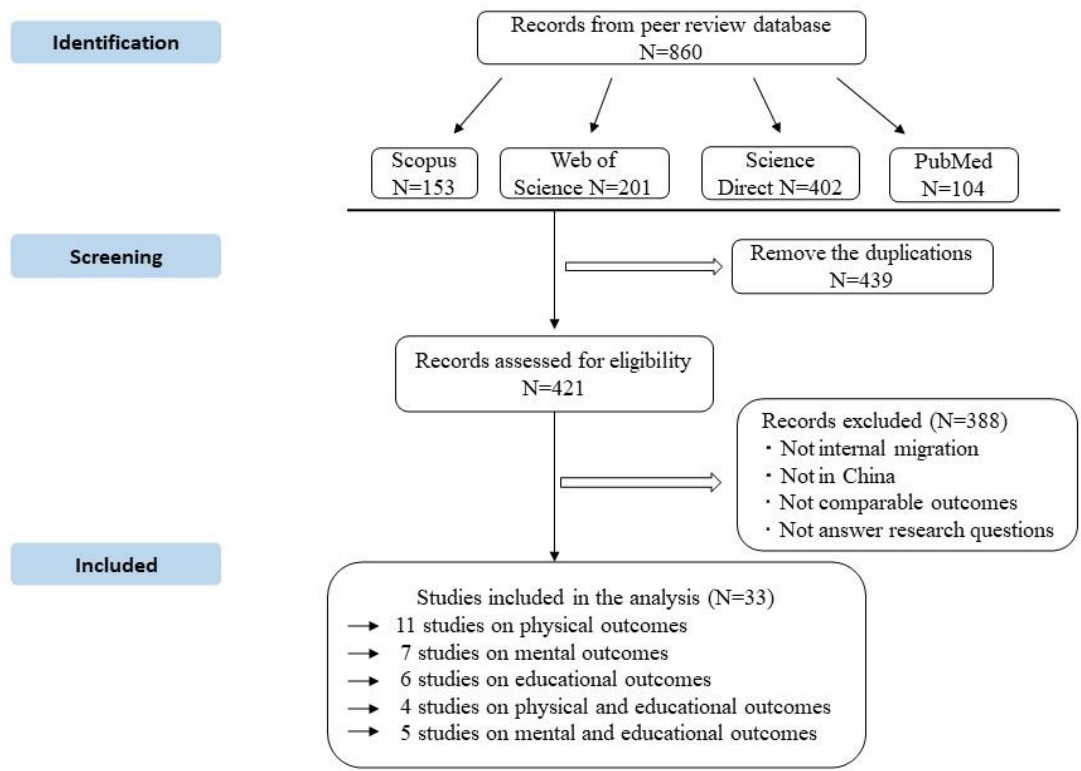

**Figure 1.** The flow chart of study selection

## 3. Results

Based on study search results from 4 indexed databases, as shown in **Figure 1**, 33 selected studies are eligible for the scoping review. In this section, we will provide detailed empirical evidence about the overall effect of parental migration on LBC's physical, mental, and educational outcomes. Considering the feasibility, to simplify the analysis and improve the accuracy, this study will try to draw accurate conclusions by prioritizing identification strategy and data quality.

### 3.1. The Analysis of LBC's Physical Outcomes

From **Table 2**, we can find that the impact of parental migration on LBC's physical outcomes is complex. A potential reason is the variability of contextual factors, such as migration pattern (who migrates and migration timing), the age and gender of the child, and the diversity of outcomes measured, which will be analyzed in detail in Section 4.2. These challenges may also occur in the following parts regarding LBC's mental and educational outcomes.

On the one hand, some evidence suggests that parental migration improves LBC's health condition, which may be explained by the channel of increased remittance from migrant parents to some extent [22]. In particular, a father's migration appears to play a more important role, and the early childhood nutrition of LBC with a father's migration performs as well as or even better than those with both parents at home in terms of stunting, underweight, and anemia risks, perhaps due to the compensating effect of the remittance from father [23].

On the other hand, some scholars draw opposite conclusions, which contradict the aforementioned opinions. LBC are prone to sickness and growth delays as a result of the lack of parental practice [24–28]. Overall, without parental supervision and care, LBC are more likely to be in poorer health [29], showing worse weight and height indices [30]. In addition, children aged 7-12 are more likely to become underweight, because LBC have to spend more time on-farm work, doing domestic work, and preparing meals [12,31] because of the lack of adult labor in the family. Noticeably, the negative impact detected in previous literature may be also biased because children's growth may have been delayed before parental migration [32]. In addition, there is also empirical evidence of no influence. Some studies found that parental migration is not associated with LBC's weight or height [32–34], childhood illness [35], and nutritional status [25].

Generally, causal identification strategies (FE, IV, PSM, and DID) combined with panel data effectively help mitigate the endogeneity issues and control for time-constant unobservables to a large extent. Parental migration exerts negative influences on LBC's Physical health. Wen et al. (2016), Li et al. (2015), and Meng et al. (2017) find the negative influences on LBC's blood pressure, illness condition, and Height/weight-for-age Z Score, which also supports this conclusion.

**Table 2.** The summary of study characteristics for LBC's physical outcomes

| Author, Year | Data Type | Child's age | Outcomes | Study design | Key Findings |
|---|---|---|---|---|---|
| Mu et al., 2015 [33] | Two-period Panel Data | 0-9 | Weight/height-for-age Z score | First-difference estimator Instrumental variable | Positive influence on LBC's weight, no influence on height |
| Wen et al., 2015 [22] | Cross-sectional Data | 10-17 | Self-rated health | Ordinary least square regression Logit regression | Positive influence |
| Shi et al., 2020 [23] | Cross-sectional Data | 6-35 months | Stunting/underweight/wasting/anemia | Generalized linear regression Logistic regression | Positive influence |
| Sun and Liang, 2021 [36] | Cross-sectional Data | 7-18 | Anemia | Inverse probability treatment weighting estimator | Positive influence with father's migration Negative influence with mother's migration |
| Wen and Li, 2016 [37] | Four-period Panel data | 7-17 | Blood pressure | Random effects estimator | Negative influence, especially with mother's or both parents' migration |
| Li et al., 2015 [27] | Cross-sectional Data | 0-18 | Illness and chronic disease | Ordinary least square regression Instrumental variable | Negative influence, especially on left-behind girls |
| Meng et al., 2017 [26] | Cross-sectional Data | 0-15 | Height/weight-for-age Z Score | Instrumental variable | Negative influence, especially on left-behind boys |
| Lu et al., 2020 [38] | Cross-sectional Data | 12-16 | Self-reported physical health | Ordered logistics regression | Negative influence |
| Lei et al., 2018 [30] | Cross-sectional Data | 1-15 | Height/weight-for-age Z Score | Ordinary least squares estimation Instrumental variable | Negative influence, particularly prominent in younger children |
| Jin et al., 2020 [29] | Cross-sectional Data | 12-17 | Self-reported Health | Ordered logistic regression | Negative influence |
| Xie et al., 2021 [35] | Panel data | 1-5 | Body mass index and illness | Hierarchical linear model | No influence |
| Xu and Xie, 2015 [39] | Cross-sectional Data | 10-15 | Self-rated health | Propensity score matching estimator | No influence |
| Tian et al., 2017 [32] | Panel data | 6-14 | Weight and height | PSM-DID estimator | No Influence |
| Zhang et al., 2015 [24] | Panel data | 0-17 | Weight and height | Growth curve models | No Influence on girls, negative influence on boys |
| Zhou et al., 2015 [34] | Cross-sectional Data | 3-10 | Height/weight-for-age Z Score | Multivariate regression analysis | No influence |

## 3.2. The Analysis of LBC's Mental Outcomes

Regarding the impact of parental migration on LBC's mental outcomes, we can draw the widely accepted conclusion of the negative impact based on the current literature in **Table 3**. Child-level mental outcomes usually consist of depression, anxiety, happiness, problem behaviors, self-esteem, and non-cognitive ability. Overall, the mental health of LBC is more susceptible to the absence of parental involvement, supervision, and interaction.

Due to the weakened support from the family environment, LBC are prone to serious mental problems. Especially compared with LBC raised by grandparents or one parent, self-guardian LBC are more disadvantaged [40,41]. More broadly, left-behind children and adolescents tend to be weak in mental adjustment [42], happiness [43], anxiety [44], loneliness, and satisfaction [45], emotional functioning, and social functioning [46], which may be caused by the lack of parent-child communication and interaction, especially both parents' or mother's accompanying [47].

As for depression, LBC are more likely to suffer from depression than those with non-migrant parents [1]. In terms of who migrates out, the mother's absence is more likely to account for children's depressive symptoms, whereas the father's absence might not play an important role [48]. As for LBC's non-cognitive ability, by comparing Big Five Index1 between LBC and non-LBC, LBC's non-cognitive ability is more vulnerable, which is also explained by the lack of parental practice, especially the parent-child interaction environment [49].

Overall, similar to former research conclusions, parental migration, may exert a negative influence on LBC's mental health and non-cognitive ability, no matter the long-term or short-term migration duration, especially for left-behind girls. This might be explained by the reason that the mother usually takes the primary responsibility for caring for children and supporting the whole family [50].

---

[1] Big Five Personality Model is an effective tool to measure child's non-cognitive ability. Big Five components mainly consist of conscientiousness, extraversion, neuroticism, agreeableness, and openness [111].

**Table 3.** The summary of study characteristics for LBC's mental outcomes

| Author, Year | Data Type | Child's age | Outcomes | Study design | Key Findings |
| --- | --- | --- | --- | --- | --- |
| Lu et al., 2020 [51] | Cross-sectional Data | 3-15 | Behavior Problems Index | Structural equation modeling | Negative influence |
| Yue et al., 2020 [52] | Cross-sectional Data | 12-13 | Depressive symptoms | Structural equation modeling | Negative influence |
| Xu et al., 2019 [48] | Two-period panel data | 13-16 | Depressive symptoms | Fixed effects model Propensity score weighting estimator | Negative influence with only mother's migration |
| Wu and Zhang, 2017 [53] | Two-period panel data | 10-13 | Non-cognitive ability | First difference estimator | Negative influence with both parents' migration |
| Zhao et al., 2017 [54] | Cross-sectional Data | 10-15 | Emotional and behavioral problems | Multiple linear regression | Negative influence |
| Zhao and Chen, 2022 [49] | Cross-sectional Data | 13-15 | Big Five Personality Index | Instrumental variable | Negative influence |
| Wu et al., 2015 [55] | Cross-sectional Data | 8-17 | Epidemiological depression score | Structural equation modeling | Negative influence |
| Shi et al., 2016 [44] | Two-period Panel Data | 12-13 | Mental health score, Social Anxiety, and Self-esteem score | Difference in difference estimator Propensity score matching estimator | Negative influence |
| Shi et al., 2020 [40] | Cross-sectional Data | 0-3 | Social-emotional problems | Structural equation modeling | Negative influence |
| Liu et al., 2021 [50] | Two-period Panel Data | 0-3 | Big Five Personality Index | Propensity score matching estimator | Negative influence with mother's long-run migration |
| Chang et al., 2019 [56] | Three-period Panel Data | 14-18 | Mental Health Test | Fixed effects model | Negative influence, especially on left behind girls |
| Wang et al., 2019 [57] | Three-period Panel Data | 13-16 | Anxiety and intrinsic motivation | Fixed effects model | Negative influence, especially on left behind girls |

### 3.3. The Analysis of LBC's Educational Outcomes

As shown in **Table 4**, the influence of parental migration on LBC's educational outcomes is complex, which may be explained by two reasons: on the one hand, due to the lack of parental supervision, LBC are usually more likely to experience grade retention or drop out of school [58–61]. On the other hand, remittance from migrant parents can increase LBC's nutritional and educational investment, which may compensate for parental absence to some extent [62–64].

At the preschool stage of LBC, children experiencing both parents' migration appear to be in a worse situation, which is mainly reflected in social skills, executive functioning, reading performance, and math test scores [65]. Therefore, the absence of both parents significantly hinders the development of an early child's cognitive ability development [35,51]. Noticeably, the mother's migration exerts an adverse influence on LBC's test scores and cognitive ability [25], much stronger than the impact of the father's migration [48]. Another point is that younger children may be more vulnerable to trauma due to parental migration than adolescents [66], illustrating that children need more parental care than adolescents when facing trouble.

At the school stage of LBC, parental migration has a negative and significant impact on academic achievement [67] and school engagement [68] of junior- or high-school adolescents. Furthermore, exam grades are adversely affected by parental migration when both parents migrate or when the left-behind parent is the principal caregiver [69]. Some scholars accurately measure the effect of parental migration on children's test scores through a field survey and find a severe negative impact in the case of both parents' migration [70,71], which is consistent with the aforementioned empirical evidence. Considering migration intensity at the community- or village- level, the more people migrate out from the community, the lower verbal and math scores and years of schooling LBC obtain [72]. In Table 5, most of the literature indicates the negative impact of parental migration on LBC's educational performance, especially for left-behind boys.

Finally, prioritizing causality identification strategy and panel data, we can get the following conclusions: First, the impact of parental migration on LBC's physical health is negative, despite the compensation effect of remittance to some extent. Second, most literature supports the point that parental migration deteriorates LBC's mental health, especially for left-behind girls. Third, for children and adolescents, there is an overall negative influence on educational performance and cognitive ability, especially for left-behind boys. Forth, both parents' migration exacerbates the deteriorating effects on LBC's development.

**Table 4.** The summary of study characteristics for LBC's educational outcomes

| Author, Year | Data Type | Child's age | Outcomes | Study design | Key Findings |
|---|---|---|---|---|---|
| Wen et al., 2015 [22] | Cross-sectional Data | 10-17 | Exam scores and educational aspiration | Ordinal logit regression | Positive influence |
| Bai et al., 2018 [73] | Two-period Panel Data | 9-12 | Standardized English test score | Difference in difference estimator; Propensity score matching estimator | Positive influence |
| Lu et al., 2020 [51] | Cross-sectional Data | 3-15 | Verbal score | Structural equation modeling | Negative influence with both parents' migration |
| Xie et al., 2019 [72] | Cross-sectional Data | 10-15 | Standardized test scores of vocabulary and math | Hierarchical linear modeling | Negative influence |
| Meng et al., 2017 [26] | Cross-sectional Data | 7-15 | Chinese and math test scores | Instrumental variable | Negative influence, especially for left-behind boys |
| Xu et al., 2019 [48] | Two-period panel data | 13-16 | Midterm test scores and cognitive test scores | Fixed effects model; Propensity score weighting estimator | Negative influence with only mother's migration |
| Zhang et al., 2014 [71] | Two-period panel data | 10-14 | Final exam score | Difference in difference estimator; Generalized method of moments | Negative influence with both parents' migration |
| Yue et al., 2020 [25] | Four-period panel data | 6-30 Months | Cognitive score | Difference in difference estimator | Negative influence with earlier migration |
| Wu and Zhang, 2017 [53] | Two-period panel data | 10-13 | Chinese and math test scores | First difference estimator | Negative influence with both parents' migration |
| Zhou et al., 2014 [69] | Cross-sectional Data | 8-17 | Chinese and math test scores | Propensity score matching estimator | Negative influence on boys left behind by two parents |
| Zhao et al., 2014 [70] | Cross-sectional Data | 11-14 | Standardized math test Score | Instrumental variable; Bivariate probit regression | Negative influence |
| Jin et al., 2020 [29] | Cross-sectional Data | 12-17 | Self-assessment academic performance | Ordered logistic regression | No influence |
| Chang et al., 2019 [56] | Three-period Panel Data | 14-18 | Mathematics score | Fixed effects model | No influence |
| Zhou et al., 2015 [34] | Cross-sectional Data | 3-10 | Test scores and dropout rate | Multivariate regression analysis | No influence |
| Wang et al., 2019 [57] | Three-period Panel Data | 13-16 | Final math test score | Fixed effects model | No influence |

# 4. Discussion

To further clarify the influence of parental migration impact on LBC, based on previous studies, we try to construct the theoretical framework as in **Figure 2** and provide a detailed analysis: (ⅰ) Mechanisms. There may be several channels to mediate or moderate this causal effect, such as the effect of remittance and parental practice, and the compensation effect of social support. (ⅱ) Heterogeneity. It is still necessary o uncover how the impact would vary in heterogeneous contextual factors, such as the pattern of parental migration and the timing of migration, parent-level education and income, and child-level gender.

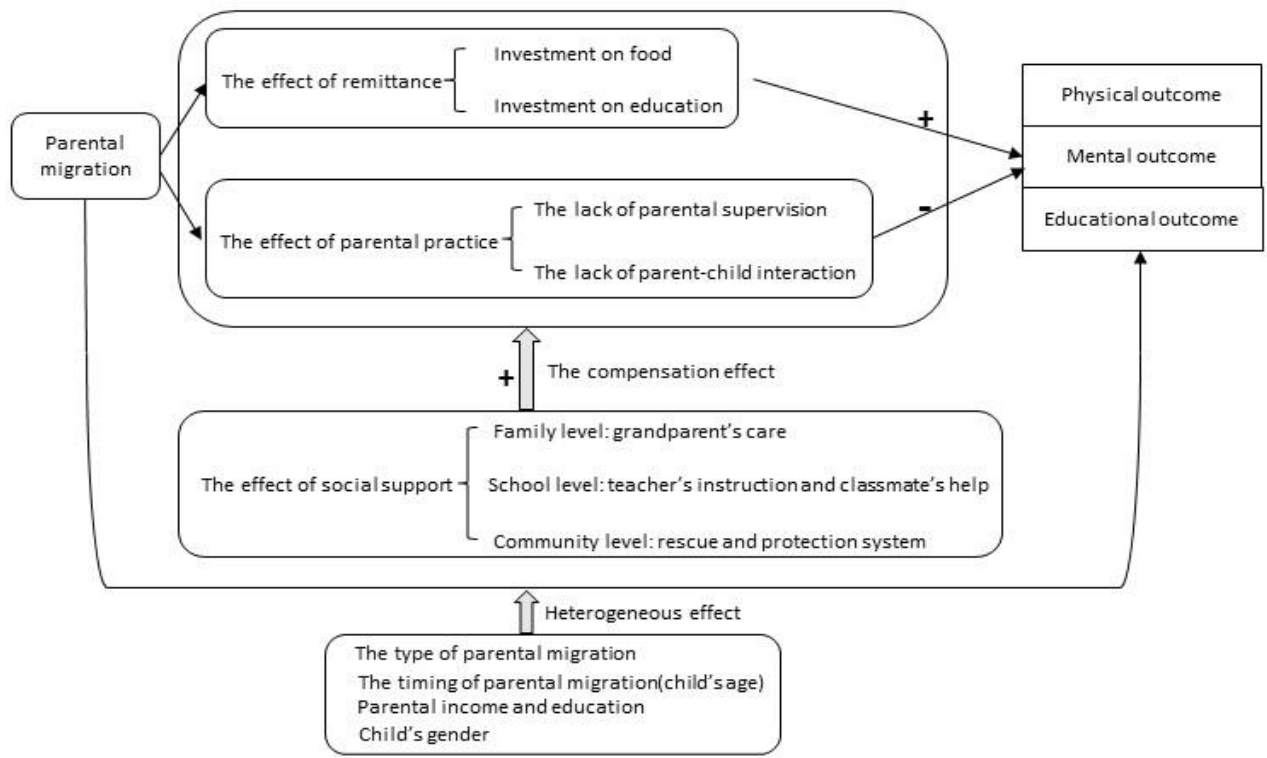

**Figure 2.** the theoretical framework between Parental Migration and LBC's Wellbeing

## 4.1 Mechanism Analysis

### 4.1.1 The Effect of Remittance

The remittance from migrant parents is beneficial to improving LBC's nutritional, educational, and mental performance and boosting their growth [73–76]. On the one hand, family income will be raised because rural adults migrate to cities to work and consequently send remittances back. With the relaxation of family budget constraints, parents are willing to invest more in children's food recipes and education [77]. In addition, migrant parents with higher education expectations for their children tend to invest more in education through remittance [73,78]. On the other hand, remittance also contributes to improving family circumstances [79], which is also essential to a child's growth. One piece of solid evidence is that migrant families are more likely to have the access to tap water compared with intact families in rural China, which protects children from viruses and bacteria to some extent [33].

### 4.1.2 The Effect of Parental Practice

The parental practice mainly includes parental supervision and parent-child communication and interaction, which is of great importance for child development [9,80]. Thus, the absence of parental practice may exert a negative impact on LBC's growth in physical health [51], and cognitive and non-cognitive ability [81,82]. Compared with father's migration [48], mother's longer migration duration and fewer interactions with children may be detrimental factors for LBC [46]. In addition, the absence of parental supervision usually leads to less parental practice in after-school tutoring children's studies [71,83], which also affects LBC studies negatively [49]. Furthermore, some scholars make a comparison between the remittance effect and the parental practice effect and conclude that the parental absence effect may outweigh the remittance effect [25].

### 4.1.3 The Effect of Social Support

The effect of social support can be deemed as the compensating effect for parental migration. Generally, social support is closely related to youth development. Children's general social needs are satisfied through social support, such as support from surrounding persons and the environment [84,85]. From the perspective of surrounding persons, the negative impact of parental absence can be compensated for to some extent by the primary caregiver in the family (usual grandparents in most cases), friends outside the family, and teachers at school [86,87]. From the environmental perspective, the well-being of LBC relies more on social support from family, school, and community [54,88]. One piece of strong evidence from a randomized controlled trial suggests that early childhood exposure to the surrounding environment is critical for children's long-term development [89,90]. In particular, family economic status and neighborhood play an important role in children's development [91]. For example, family-level and community-level social support can help mitigate the negative effect of parental absence to some extent [55].

## 4.2 Heterogenous Analysis

### 4.2.1 The Influence of Who Migrates: Father, Mother, or Both Parents

Parental migration is generally divided into three types: only father migrating, only mother migrating, and both parents migrating. There are significant differences between these three patterns. First, existing literature has verified that both parents' absence will exert more severe negative impacts on LBC's mental and educational performance [48,71]. Both parents' accompanying and supervision are of great importance for a child's growth [92]. Due to the lack of both parents' accompanying and caring, the detrimental effect on LBC's mental well-being may be amplified to a large extent [45,92]. Second, mother's absence may hinder children's development more severely than father's absence [25,36,42]. One possible explanation is that maternal care and parent-child interaction play a vital role in a child's brain development [93,94]. Another possible explanation is that father's larger remittance effect may have offset the negative impact of father's migration because there is gender discrimination in the Chinese labor market. Male workers can usually earn a higher salary than female workers [95], so father is usually the first choice to migrate out for work.

**4.2.2 The Timing of Migration in Early Childhood**

Migration timing is another crucial factor. From the perspective of the cumulative advantage/disadvantage theory, one's experience in the life trajectory is a cumulative process based on the (dis)advantage at the initial stage [96,97]. Previous literature has also provided empirical evidence that the long-term impact of parental migration on LBC's health will be particularly prominent for LBC whose parents migrated in their early childhood [30,98]. More specifically, compared with parental migration taking place at a later age stage, the causal effect of migration on children aged 24–30 months may be more detrimental [25]. Furthermore, if the timing of parental migration is early for children, the duration of migration is also a point worthy of note. Usually, the longer the parental migration duration, the greater the negative influence on LBC's development [26]. However, there is opposite evidence that the accumulated years of parental migration do not affect the schooling of the LBC [77], which may be caused by other mechanisms, such as enough remittance or grandparents' meticulous care.

**4.2.3 Parental Income and Education**

Parental education and income are usually deemed as important parental-level or family-level factors. It has been widely accepted that family socioeconomic status usually exerts profound effects on a child's development [99]. From the perspective of family income, there are usually two explanations: First, higher family income means a higher household budget constraint, which could diminish the motivation of parents to migrate out for work [100]. Thus, parents can spend more time with their children, beneficial to children's development. Second, in families with higher income, children can obtain better care and education, such as nutritious food, good-quality early education training, and an excellent family environment, greatly stimulating children's cognitive and non-cognitive development [101,102]. From the perspective of parental education, highly educated parents may tend to have more educated grandparents who can care for the grandchildren. Thus, highly educated grandparents may have a better capacity to supervise LBC to compensate for the negative influence of parental absence [103]. In addition, parents with higher education are more likely to have higher educational expectations for their children, so they are willing to invest more in their children's studies [104].

**4.2.4 The Gender of Left-behind Children**

LBC's gender is also one important issue [88]. At present, scholars don't draw definite conclusions about either left-behind boys or girls are more likely to be disadvantaged, mainly because of the variation in LBC's outcomes. From the perspective of physical health, girls are more vulnerable than boys in height and weight, and the disadvantage will be amplified if parental migration continues for a longer duration [27,32]. From the perspective of daily behaviors, compared with left-behind girls, left-behind boys are at higher risk of smoking, computer game addiction, skipping breakfast, and suicide willingness without parental supervision and regulation [43]. From the perspective of mental health, left-behind girls are more likely to suffer from severe mental problems than boys, such as depression symptoms and anxiety [41,57]. In terms of educational outcomes, empirical evidence is that both parents' migrating out only adversely affects boys' educational performance, but not girls' [69], which may be explained by the fact that girls may have more patience and persistence in studies.

## 5. Conclusions and Implications

### 5.1 Conclusions

Within the Arksey and O'Malley framework, this scoping review analyzed 33 research articles and collected empirical evidence that identified the impact of parental migration on LBC's physical, mental and educational outcomes in detail. Parental migration may deteriorate LBC's well-being from various aspects of child development, regardless of children's age, gender, and the pattern of parental migration. It is worth noting that both parents' migration may exacerbate the situation of LBC. In terms of physical outcome, LBC will have poorer health conditions than children accompanied by both parents, despite the compensatory role of remittances. In terms of mental outcome, LBC are more vulnerable to mental problems and non-cognitive development delays, especially for left-behind girls. In terms of educational outcome, there is a negative impact on LBC's cognitive abilities and school performance, especially for left-behind boys. These conclusions are enlightening and urge that more attention should be paid to LBC's development without parental care and supervision. Ensuring LBC's development and human capital accumulation is of great importance for achieving *Sustainable Development Goal 3* (Good health and Well-being) and *Sustainable Development Goal 10* (Reduced Inequalities) [3,4]. Thus, it is imperative to grant full access to the range of public and financial services for caring for LBC, thereby improving LBC's well-being and mitigating inequalities between migrants' and non-migrants' children.

## 5.2 Implications

Based on the above research, this study tries to present 3 implications, which may be helpful to fill the gaps for future research on LBC. First, researchers and policymakers should consider how to construct large-scale representative longitudinal survey data within the context of macro-level *Hukou* reforms in China. This large-scale longitudinal survey should include migrants and original family members simultaneously, especially focusing on collecting the well-documented records of migration, such as migration timing, migration duration, and the experience of migrating out and back. Rich data will offer researchers a wider foundation to understand the pattern of migration and relative mechanisms [105], which can increase the credibility and availability of causal identification to some extent.

Second, try to find effective IVs, such as the policy regarding migration, to solve the endogeneity problems. Migration decisions can be endogenous to the outcomes (even after accounting for individual fixed effects) [26,70,71]. The aforementioned instrumental variables are retrieved mainly from climate change, migration networks, and economic growth. Notably, it is worthwhile to learn from a few studies that have constructed quasi-natural experiments. For example, one unique policy, called Pacific Access Category (PAC), introduced immigration-visa lotteries for randomly selecting applicants from Tongan. This identification strategy masterfully solves selection bias because participants both in treated and control groups have an equal willingness to migrate [106–108]. However, it seems to be challenging to find this kind of quasi-natural experiment, which mainly relies on the cost of implementing such a policy and the willingness of the policymakers.

Third, it is also important to pay more attention to *Hukou* reform in China, left-behind children, and migrant children simultaneously. Given the context of loosening restrictions on *Hukou* conversion, migrant workers now have easier access to *Hukou* in cities [109]. In recent years, there has been an important trend that the number of migrant children is increasing year by year. The well-being of migrant children in urban areas is also worthy of attention with the same importance as LBC. It is still confusing which is the better decision for migrant parents: to leave their children behind or take them into cities together. Furthermore, with the increase in migrant children and the slight decrease in LBC [110], it remains unclear whether LBC and migrant children can obtain sufficient social support or not. How to support two types of children in migrant families need more reflection from researchers and policymakers.

**Funding:** This research was funded by China Scholarship Council, grant number 202003250063.

**Data Availability Statement:** Not applicable.

**Acknowledgments:** The author would like to be grateful for the precious comments from Professor Dr. Bart Cockx from Ghent University. The funding support is from China Scholarship Council (202003250063) acknowledged by the author.

**Conflicts of Interest:** The authors declare no conflict of interest.